\documentclass{article}

\usepackage{PRIMEarxiv}
\usepackage[utf8]{inputenc} 
\usepackage[T1]{fontenc}    
\usepackage{hyperref}       
\usepackage{url}            
\usepackage{booktabs}       
\usepackage{amsfonts}       
\usepackage{nicefrac}       
\usepackage{microtype}      
\usepackage{lipsum}
\usepackage{fancyhdr}       
\usepackage{graphicx}       
\graphicspath{{media/}}     
\usepackage{caption}
\usepackage{subcaption}
\usepackage{float}
\usepackage{url}
\usepackage{amsmath}
\usepackage[numbers]{natbib}

\newcommand\numberthis{\addtocounter{equation}{1}\tag{\theequation}}

\title{Multimodal Intrinsic Speckle-Tracking (MIST) to extract rapidly-varying diffuse X-ray scatter}

\author{
  Samantha J Alloo*,\\
  School of Physical and Chemical Sciences, University of Canterbury, Christchurch, New Zealand \\
  \texttt{samantha.alloo@pg.canterbury.ac.nz} \\
   \And
  Kaye S Morgan, \\
  School of Physics and Astronomy, Monash University, Victoria, Australia\\
   \And
   David M Paganin,\\
   School of Physics and Astronomy, Monash University, Victoria, Australia\\
   \And 
   Konstantin M Pavlov,\\
   School of Physical and Chemical Sciences, University of Canterbury, Christchurch, New Zealand. \\School of Physics and Astronomy, Monash University, Victoria, Australia.\\School of Science and Technology, University of New England, Armidale, Australia \\
}

\begin{document}
\maketitle

\begin{abstract}
Speckle-based phase-contrast X-ray imaging (SB-PCXI) can reconstruct high-resolution images of weakly-attenuating materials that would otherwise be indistinguishable in conventional attenuation-based imaging. The experimental setup of SB-PCXI requires only a sufficiently coherent source and spatially random mask, positioned between the source and detector. The technique can extract sample information at length scales smaller than the imaging system's spatial resolution; this enables multimodal signal reconstruction. ``Multimodal Intrinsic Speckle-Tracking'' (MIST) is a rapid and deterministic formalism derived from the paraxial-optics form of the Fokker-Planck equation. MIST simultaneously extracts attenuation, refraction, and small-angle scattering (diffusive-dark-field) signals from a sample and is more computationally efficient compared to alternative speckle-tracking approaches. Hitherto, variants of MIST have assumed the diffusive-dark-field signal to be spatially slowly varying. Although successful, these approaches have been unable to well-describe unresolved sample microstructure whose statistical form is not spatially slowly varying. Here, we extend the MIST formalism such that there is no such restriction, in terms of a sample's rotationally-isotropic diffusive-dark-field signal. We reconstruct multimodal signals of two samples, each with distinct X-ray attenuation and scattering properties. The reconstructed diffusive-dark-field signals have superior image quality compared to our previous approaches which assume the diffusive-dark-field to be a slowly varying function of transverse position. Our generalisation may assist increased adoption of SB-PCXI in applications such as engineering and biomedical disciplines, forestry, and palaeontology, and is anticipated to aid the development of speckle-based diffusive-dark-field tensor tomography.
\end{abstract}


\section{Introduction}
\begin{figure}[tb]
    \centering
    \includegraphics[width=0.6 \linewidth]{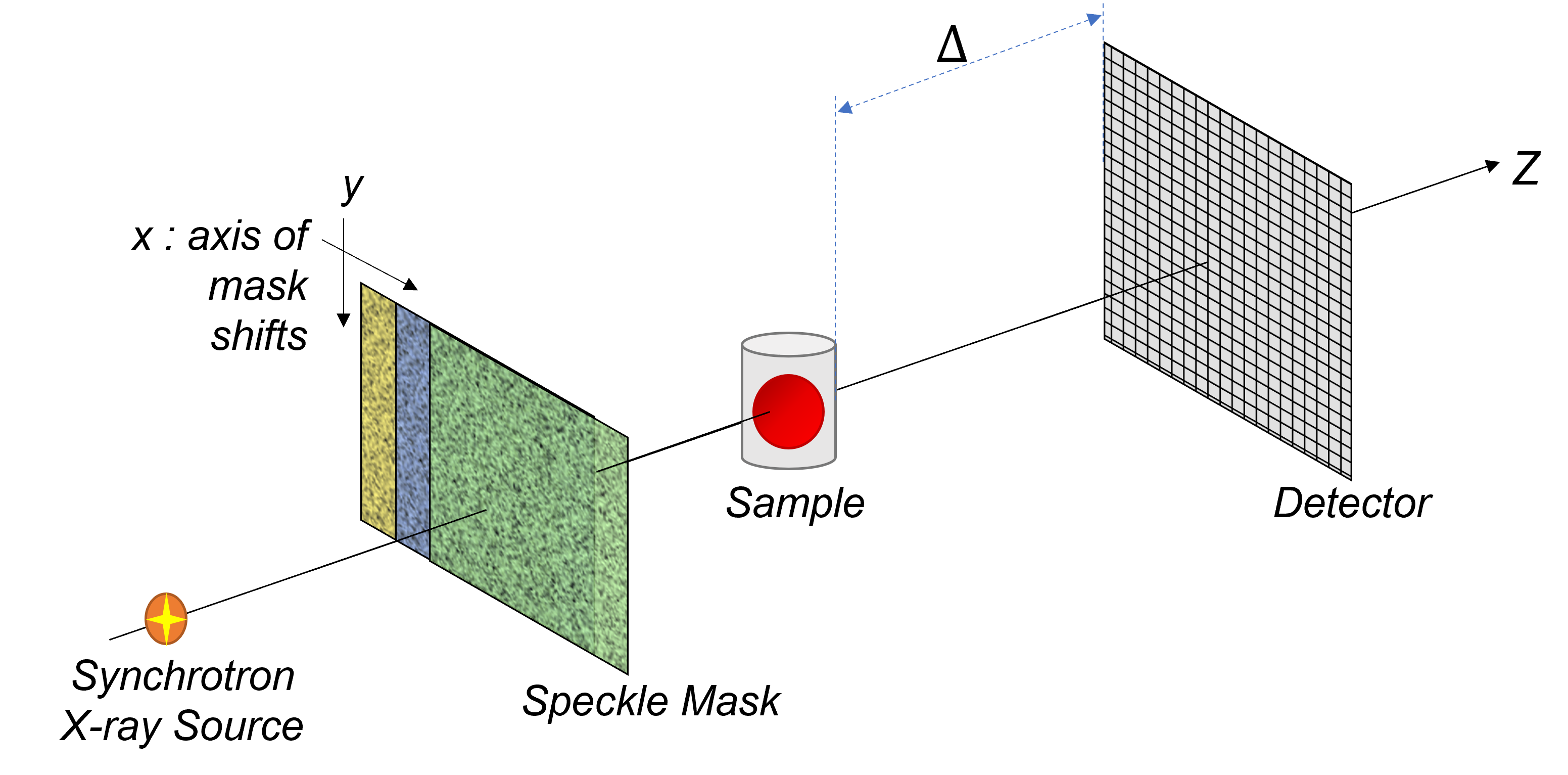}
    \caption{\textit{Experimental setup of speckle-based phase-contrast X-ray imaging using a synchrotron X-ray source.}}
    \label{fig:setup}
\end{figure}

Phase-contrast X-ray imaging (PCXI)\cite{paganin2006coherent} techniques can reconstruct high-resolution images of weakly-attenuating materials that would otherwise be indistinguishable in conventional attenuation-based imaging, for example, resolving soft-tissues in conventional radiography. PCXI techniques utilise X-ray phase-shifts and attenuation, described by the material's real and imaginary components of its refractive index,
${n({\mathbf{r}}')=1 - \delta({\mathbf{r}}') + i \beta({\mathbf{r}}')}$, 
respectively, to reconstruct images. Here ${\mathbf{r}}'$ is the three-dimensional position vector.  This is a vital point of difference compared to attenuation-based imaging, which only exploits X-ray attenuation. The most common approach to converting sample-imposed phase effects into measurable detector intensity differences is to introduce sophisticated optical elements into the experimental setup, for example, via grating-interferometry\cite{momose2003demonstration}, analyser-based imaging\cite{forster1980double}, grid-based imaging\cite{wen2008spatial,morgan2011quantitative}, and edge-illumination imaging\cite{olivo2001innovative}. To eliminate the need for complex optics, propagation-based\cite{snigirev1995possibilities, cloetens1996} PCXI has also been developed, however, it does require sufficiently large sample-to-detector distances and has stringent coherence requirements. The superiority of PCXI techniques over attenuation-based imaging has already been proven in multiple disciplines. Such studies have been performed using synchrotron and conventional laboratory X-ray sources for applications in several fields, including biomedical\cite{jung2022mucociliary,drevet2022new,sena2022synchrotron,massimi2019laboratory}, agricultural and food sciences\cite{indore2022synchrotron}, palaeontology\cite{mcneil2010imaging,edgecombe2012scolopocryptopid}, and materials science\cite{denecke2011speciation,mayo2006laboratory}. 
\\\\
PCXI techniques can also provide information regarding sample structures at length scales smaller than the spatial resolution of the imaging system, namely, through small-angle X-ray scattering (SAXS) (or diffusive-dark-field (DF) imaging) from unresolved sample microstructure. PCXI techniques that reconstruct the coherent, phase-contrast (PC), and diffuse, DF, flows of an X-ray wavefield are often termed ``multimodal'', in the sense that they can extract multiple signals. Grating interferometry\cite{pfeiffer2008hard}, analyser-based imaging\cite{pagot2003method,wernick2003multiple}, grid-based imaging\cite{how2022quantifying}, edge-illumination imaging\cite{endrizzi2015edge}, and propagation-based imaging\cite{gureyev2020dark,leatham2021x} are examples of PCXI techniques that are sensitive to SAXS, and can therefore reconstruct a sample's DF signal.
\\\\
Speckle-based (SB) PCXI, first developed in 2012\cite{berujon2012two,morgan2012x}, is a particularly appealing multimodal technique as it is experimentally simple, cost-effective, and radiation dose-efficient. A speckle pattern is generated when a sufficiently coherent wavefield propagates through a membrane with random refractive index fluctuations\cite{goodman2020speckle}. A speckle pattern, in the context of SB-PCXI, is used as an X-ray wavefront marker whose subsequent modification is used to measure sample-induced speckle modulations, e.g., transverse spatial shifts, attenuation, and blurring. The SB-PCXI experimental setup consists of an X-ray source,  speckle-generating mask, sample, and detector system, positioned some finite distance downstream from the sample, as shown in Fig.~\ref{fig:setup}. In this work, we consider a synchrotron X-ray source that produces a monochromatic, paraxial wavefield having a high degree of both spatial and temporal coherence. The X-ray wavefield is randomly modulated by propagation through the speckle-generating mask (e.g., conventional sandpaper) and is then registered by the position-sensitive detector positioned downstream. We highlight two particularly attractive features of SB-PCXI. Firstly, the speckle-generating mask can be any spatially-random medium.  This removes the restriction of needing precisely-manufactured optical elements, making the experimental setup both easy to implement and flexible. Secondly, SB-PCXI requires relatively low spatial and temporal coherence\cite{zdora2018state}.
\\\\
The inverse problem\cite{Sabatier2000} of SB-PCXI involves reconstructing a sample's multimodal signals, given suitable reference-speckle and sample-reference-speckle intensity images. The reference-speckle images resemble the composition of the mask and the sample-reference-speckle images are captured when a sample is placed into this reference-speckle field. This reference-speckle pattern is modified, depending on the refractive properties of the sample, and these speckle modifications are used to reconstruct sample information. Transverse speckle shifts are associated with the PC signal, whereas speckle blurring, or reduction in visibility, is correlated to the sample's DF signal. There are two distinct approaches in the research literature to solve the multimodal inverse SB-PCXI problem, namely, extrinsic and intrinsic approaches. X-ray speckle-vector tracking\cite{berujon2012two} (XSVT), mixed XSVT approaches\cite{berujon2016x,berujon2017near}, X-ray speckle-scanning\cite{wang2016synchrotron,wang2016high} (XSS), and unified modulated pattern analysis\cite{zdora2017x} (UMPA) are examples of extrinsic approaches that reconstruct multimodal signals using iterative pixel-wise methods on SB-PCXI data acquired at multiple random-mask positions. In essence, XSVT and mixed XSVT approaches perform a zero-normalised cross-correlation analysis, between the reference-speckle and sample-reference-speckle images. UMPA proposes a different methodology, based on least-squares minimisation between a model and the measurement of the sample-reference-speckle pattern across all mask positions. These techniques have the advantage of a relatively short image acquisition time compared to XSS\cite{wang2016synchrotron,wang2016high} techniques. The SB-PCXI inverse problem was reconceptualised in 2018 when Paganin \textit{et al.}\cite{paganin2018single} proposed a geometric-flow approach to reconstruct a sample's PC signal using a single set of SB-PCXI intensity data; this was the first realisation of so-called intrinsic speckle-tracking, since it does not explicitly track individual speckles, but rather solves a partial differential equation formulated at the whole-of-image level. This speckle-tracking geometric-flow formalism\cite{paganin2018single} was then combined with a Fokker-Planck-type\cite{paganin2019x,morgan2019applying} generalisation of the transport-of-intensity equation\cite{Teague1983} of paraxial wave optics to allow for multimodal intrinsic signal extraction\cite{pavlov2020x}, which was named ``Multimodal Intrinsic Speckle-Tracking'' (MIST). The Fokker-Planck\cite{risken1996fokker} expression is based on local energy conservation, and it considers transverse radiation flows as a combination of coherent and diffusive effects; these effects for the case of multimodal signal extraction align with the PC and DF signals, respectively. MIST is less computationally expensive than the alternative extrinsic approaches\cite{berujon2012two,morgan2012x,zdora2017x,berujon2016x,berujon2017near,wang2016synchrotron,wang2016high}. This means that MIST can be used to rapidly reconstruct tomographic data\cite{alloo2022dark} of both the PC and DF signals, therefore, to provide three-dimensional sample information that is inaccessible in single-projection imaging.
\\\\
MIST was first developed\cite{pavlov2020x} under three key assumptions: (a) the sample is a pure phase-object, such that X-ray attenuation can be neglected, (b) the unresolved sample microstructure diffusely scatters the X-ray beam in a rotationally-isotropic manner, and (c) the sample's DF signal is a slowly varying function of transverse position. The first assumption, (a), was relaxed in our most recent work where we presented the case of attenuating materials having rotationally-isotropic position-dependent diffuse scatter\cite{alloo2022dark}. Assumption (b) was relaxed when we considered anisotropically-scattering attenuating materials\cite{pavlov2021directional}, such that directional DF
\cite{jensen2010a,jensen2010b} signals could be obtained. Here, we should highlight that we have also performed DF computed-tomography (CT) using MIST\cite{alloo2022dark}, as the reconstructed two-dimensional DF signals had a sufficiently high signal-to-noise ratio (SNR) and spatial resolution such that they could be used in standard CT reconstruction algorithms, such as filtered back-projection\cite{kak2001principles}. Assumption (c), which considers the sample's DF signal to be spatially slowly varying, is a condition that has remained in all of the MIST approaches to date. Although both Pavlov \textit{et al.}\cite{pavlov2020x} and Alloo \textit{et al.}\cite{alloo2022dark} have demonstrated that this approach to MIST is capable of reconstructing images with a high spatial resolution, which is at least comparable to the alternative extrinsic approaches, our recent investigations have found that this approximation may break down at sharp  interfaces.  This assumption may hinder the full potential of previous MIST approaches. In particular, samples that have rapidly varying microstructure-autocorrelation functions may breach the domain of applicability of all previous MIST approaches. In the present paper, we generalise MIST to alleviate this restriction, thereby broadening its domain of utility. Several studies have verified the broad applicability and importance of DF imaging\cite{miller2013phase,wang2016high,lim2022low}, with a significant focus on biomedical clinical applications\cite{ando2016dark}, e.g., using DF imaging for early-stage diagnosis of lung diseases such as fibrosis\cite{yaroshenko2015improved,hellbach2017x}, pneumothorax\cite{schleede2012emphysema}, emphysema\cite{meinel2013diagnosing}, and breast cancer\cite{gureyev2020dark}. Our improved MIST approach could provide an alternative experimentally versatile, low-dose imaging technique that can reconstruct high-resolution multimodal signals in two- and three-dimensions. 
\\\\
This paper progresses as follows. First, we theoretically develop the new generalised MIST approach, deriving an analytical solution for a sample's phase-shift (PC signal) and effective diffusion coefficient (DF signal). Numerical stabilisation techniques are then discussed, before applying the approach to synchrotron SB-PCXI data of two samples that have different X-ray attenuation and SAXS characteristics. Moreover, we first consider a sample that is weakly-attenuating (almost a pure phase-object), and then a more attenuating object to investigate the breadth of applicability of our approach. We then compare this new approach, qualitatively and quantitatively, to both of the published rotationally-isotropic MIST approaches\cite{pavlov2020x,alloo2022dark}: (a) Pavlov \textit{et al.}'s approach\cite{pavlov2020x}, which neglects X-ray attenuation, and (b) Alloo \textit{et al.}'s approach\cite{alloo2022dark}, which considers it.  Note that both previous approaches approximate the DF to be spatially slowly varying. The paper is finished by discussing potential future research avenues. 

\section*{Theoretical derivation of updated MIST approach}

The SB-PCXI form of the Fokker-Planck equation models the coherent and diffusive flows of X-rays in the case of a phase-object, as described by its phase-shift, $\phi_{\textrm{ob}}(\textbf{r})$, and effective diffusion coefficient, $D_{\textrm{eff, Phase}}(\textbf{r};\Delta)$, respectively. The SB-PCXI Fokker-Planck equation is derived assuming the reference-speckle field is spatially well-resolved, and that this field, alongside the speckle field in the presence of the sample, obeys a Fokker-Planck\cite{risken1996fokker} extension of the geometric-flow formalism for speckle tracking\cite{paganin2018single}. The Fokker-Planck equation in this instance is\cite{paganin2019x}  
\begin{equation}
\label{eqn:FPPhase}
I_R(\textbf{r})-I_S(\textbf{r}) = \frac{\Delta}{k}\nabla_{\perp}\cdot \left[I_{R}(\textbf{r})\nabla_{\perp}\phi_{\textrm{ob}}(\textbf{r})\right]-\Delta\nabla_{\perp}^{2}\left[D_{\textrm{eff, Phase}}(\textbf{r};\Delta)I_{R}(\textbf{r})\right],
\end{equation}
where $I_R(\textbf{r})$ and $I_S(\textbf{r})$ denote the reference-speckle and sample-reference-speckle intensities, respectively, $\textbf{r}\equiv(x,y)$ denotes Cartesian coordinates in planes perpendicular to the optical axis $z$, assuming plane-wave illumination, $\Delta$ is the sample-to-detector distance, $k$ is the wavenumber, $\nabla_{\perp} = (\partial/\partial x, \partial/\partial y)$ is the transverse gradient operator in the $(x,y)$ plane, and  $\nabla_{\perp}^{2}$ is the transverse Laplacian operator. The reference-speckle intensity, $I_R(\textbf{r})$, is obtained with the speckle mask in and the object out of the X-ray beam, and the sample-reference-speckle intensity, $I_S(\textbf{r})$, is obtained by placing a phase- and SAXS-inducing object into the speckle-modulated X-ray beam. In previous rotationally-isotropic MIST approaches\cite{pavlov2020x,alloo2022dark}, the Fokker-Planck equation has been simplified (for phase-objects) to a linear equation in terms of the effective diffusion coefficient and the Laplacian of the phase-shift. This simplification resulted by assuming the effective diffusion coefficient to be spatially slowly varying, such that various second-order terms could be neglected. In the present work, we do not make this assumption, and instead, the effective diffusion coefficient can accurately describe the diffuse scattering signal from unresolved sample microstructure for which the autocorrelation function can be a rapidly varying function of transverse position.
\\\\
The present approach begins by expanding the coherent flow term, namely the first term on the right side of the preceding equation, into its two components that describe the lensing and prism-like effects\cite{paganin2019x}:
\begin{equation}
\frac{\Delta}{k}\nabla_{\perp}\cdot \left[I_{R}(\textbf{r})\nabla_{\perp}\phi_{\textrm{ob}}(\textbf{r})\right] = \frac{\Delta}{k}\left[I_{R}(\textbf{r})\nabla_{\perp}^2\phi_{\textrm{ob}}(\textbf{r}) + \nabla_{\perp}I_{R}(\textbf{r})\cdot\nabla_{\perp}\phi_{\textrm{ob}}(\textbf{r})\right].
\end{equation}
Following the approximation described in Pavlov \textit{et al.}\cite{pavlov2020x}, we neglect the scalar product of the gradient of the random rapidly varying wavefield intensity $I_{R}(\textbf{r})$ with the more slowly changing gradient of the wavefield phase $\phi_{\textrm{ob}}(\textbf{r})$, namely, we assume that 
\begin{equation}
\frac{\Delta}{k}\nabla_{\perp}\cdot \left[I_{R}(\textbf{r})\nabla_{\perp}\phi_{\textrm{ob}}(\textbf{r})\right] \approx \frac{\Delta}{k} I_{R}(\textbf{r})\nabla_{\perp}^2\phi_{\textrm{ob}}(\textbf{r}).
\end{equation}
Hence equation~(\ref{eqn:FPPhase}) becomes (cf.~Ref.~\citenum{pavlov2020x}) 
\begin{equation}
\label{eqn:FPPhaseSim}
I_R(\textbf{r})-I_S(\textbf{r}) = \frac{\Delta}{k}I_{R}(\textbf{r})\nabla_{\perp}^2\phi_{\textrm{ob}}(\textbf{r})-\Delta\nabla_{\perp}^{2}\left[D_{\textrm{eff, Phase}}(\textbf{r};\Delta)I_{R}(\textbf{r})\right].
\end{equation}
The associated inverse problem of multimodal SB-PCXI, in this work, involves solving equation~(\ref{eqn:FPPhaseSim}) for $D_{\textrm{eff, Phase}}(\textbf{r};\Delta)$ and $\phi_{\textrm{ob}}(\textbf{r})$, given the SB-PCXI intensity data, $I_R(\textbf{r})$ and $I_S(\textbf{r})$. We expand the final term on the right side of equation~(\ref{eqn:FPPhaseSim}), to give the following alternative, but still equivalent, form of a phase-object's SB-PCXI Fokker-Planck expression: 
\begin{align}
  \label{eqn:FPPhaseSimEx}
\frac{1}{\Delta}\left[I_R(\textbf{r})-I_S(\textbf{r})\right] = I_{R}(\textbf{r})\nabla_{\perp}^2\left[\frac{1}{k}\phi_{\textrm{ob}}(\textbf{r})-D_{\textrm{eff, Phase}}(\textbf{r};\Delta)\right]-D_{\textrm{eff, Phase}}(\textbf{r};\Delta)\nabla_{\perp}^{2}I_{R}(\textbf{r})\\\nonumber-2D_{\textrm{eff, Phase}}^x(\textbf{r};\Delta)I_{R}^x(\textbf{r})-2D_{\textrm{eff, Phase}}^y(\textbf{r};\Delta)I_{R}^y(\textbf{r}).  
\end{align}
This is a linear equation in terms of four unknowns---namely, $\nabla_{\perp}^2\left[\frac{1}{k}\phi_{\textrm{ob}}(\textbf{r})-D_{\textrm{eff, Phase}}(\textbf{r};\Delta)\right]$, $D_{\textrm{eff, Phase}}(\textbf{r};\Delta)$, $D_{\textrm{eff, Phase}}^x(\textbf{r};\Delta)$, and $D_{\textrm{eff, Phase}}^y(\textbf{r};\Delta)$---where the partial derivatives in the spatial coordinates \textit{x} and \textit{y} are denoted with superscripts \textit{x} and \textit{y}, respectively. These four functions can then be employed to reconstruct the sample's true phase-shifts and effective diffusion coefficient. To obtain unique solutions for four unknown variables we require four equations; to do this, four unique forms of equation~(\ref{eqn:FPPhaseSimEx}) can be generated by taking four independent measurements of $I_R(\textbf{r})$ and $I_S(\textbf{r})$. This can be achieved by, for example, transversely shifting the mask to generate a new reference-speckle pattern. Four independent measurements, denoted with subscripts $a, b$, $c$, and $d$, will induce coherent and diffuse X-ray flows that satisfy the following system of equations:
\begin{align*}
\label{eqn:FPPhaseCoupleSysa}
\frac{1}{\Delta}\left[I_{R_{a}}(\textbf{r})-I_{S_{a}}(\textbf{r})\right] = I_{R_{a}}(\textbf{r})\nabla_{\perp}^2\left[\frac{1}{k}\phi_{\textrm{ob}}(\textbf{r})-D_{\textrm{eff, Phase}}(\textbf{r};\Delta)\right]-D_{\textrm{eff, Phase}}(\textbf{r};\Delta)\nabla_{\perp}^{2}I_{R_{a}}(\textbf{r})\\-2D_{\textrm{eff, Phase}}^x(\textbf{r};\Delta)I_{R_{a}}^x(\textbf{r})-2D_{\textrm{eff, Phase}}^y(\textbf{r};\Delta)I_{R_{a}}^y(\textbf{r}),\numberthis\\
\label{eqn:FPPhaseCoupleSysb}
\frac{1}{\Delta}\left[I_{R_{b}}(\textbf{r})-I_{S_{b}}(\textbf{r})\right] = I_{R_{b}}(\textbf{r})\nabla_{\perp}^2\left[\frac{1}{k}\phi_{\textrm{ob}}(\textbf{r})-D_{\textrm{eff, Phase}}(\textbf{r};\Delta)\right]-D_{\textrm{eff, Phase}}(\textbf{r};\Delta)\nabla_{\perp}^{2}I_{R_{b}}(\textbf{r})\\-2D_{\textrm{eff, Phase}}^x(\textbf{r};\Delta)I_{R_{b}}^x(\textbf{r})-2D_{\textrm{eff, Phase}}^y(\textbf{r};\Delta)I_{R_{b}}^y(\textbf{r}),\numberthis\\
\label{eqn:FPPhaseCoupleSysc}
\frac{1}{\Delta}\left[I_{R_{c}}(\textbf{r})-I_{S_{c}}(\textbf{r})\right] = I_{R_{c}}(\textbf{r})\nabla_{\perp}^2\left[\frac{1}{k}\phi_{\textrm{ob}}(\textbf{r})-D_{\textrm{eff, Phase}}(\textbf{r};\Delta)\right]-D_{\textrm{eff, Phase}}(\textbf{r};\Delta)\nabla_{\perp}^{2}I_{R_{c}}(\textbf{r})\\-2D_{\textrm{eff, Phase}}^x(\textbf{r};\Delta)I_{R_{c}}^x(\textbf{r})-2D_{\textrm{eff, Phase}}^y(\textbf{r};\Delta)I_{R_{c}}^y(\textbf{r}),\numberthis\\
\label{eqn:FPPhaseCoupleSysd}
\frac{1}{\Delta}\left[I_{R_{d}}(\textbf{r})-I_{S_{d}}(\textbf{r})\right] = I_{R_{d}}(\textbf{r})\nabla_{\perp}^2\left[\frac{1}{k}\phi_{\textrm{ob}}(\textbf{r})-D_{\textrm{eff, Phase}}(\textbf{r};\Delta)\right]-D_{\textrm{eff, Phase}}(\textbf{r};\Delta)\nabla_{\perp}^{2}I_{R_{d}}(\textbf{r})\\-2D_{\textrm{eff, Phase}}^x(\textbf{r};\Delta)I_{R_{d}}^x(\textbf{r})-2D_{\textrm{eff, Phase}}^y(\textbf{r};\Delta)I_{R_{d}}^y(\textbf{r}).\numberthis
\end{align*}
The phase-shift $\phi_{\textrm{ob}}(\textbf{r})$, together with the effective diffusion coefficient $D_{\textrm{eff, Phase}}(\textbf{r};\Delta)$ and its transverse derivatives, are exclusively defined by the object and are hence equal for equations~(\ref{eqn:FPPhaseCoupleSysa})\textendash(\ref{eqn:FPPhaseCoupleSysd}). 
Equations~(\ref{eqn:FPPhaseCoupleSysa})\textendash(\ref{eqn:FPPhaseCoupleSysd}) can be solved using, for example, a standard Gaussian-elimination. Here, it is important to highlight that previous MIST approaches for rotationally-isotropic diffuse scatter \cite{pavlov2020x,alloo2022dark} require SB-PCXI data from just two mask positions. Within the present method, we require four mask positions since we extract two additional quantities, namely $D_{\textrm{eff, Phase}}^x(\textbf{r};\Delta)$ and $D_{\textrm{eff, Phase}}^y(\textbf{r};\Delta)$. These terms are small for samples whose unresolved microstructure can be considered to have an autocorrelation that is a slowly varying function of transverse position. However, these terms are significant in the contrary case where the unresolved sample-microstructure-autocorrelation is rapidly varying. These terms are also important to correctly reconstruct sample edges. 
\\\\
A sample's true effective diffusion coefficient can then be reconstructed by aggregating the three extracted quantities, $D_{\textrm{eff, Phase}}(\textbf{r};\Delta)$, $D_{\textrm{eff, Phase}}^x(\textbf{r};\Delta)$, and $D_{\textrm{eff, Phase}}^y(\textbf{r};\Delta)$. We employ a technique adopted in various differential imaging techniques\cite{arnison2004linear,kottler2007two,de2008quantitative,morgan2011quantitative} whereby a two-dimensional function, $g(x,y)$, can be calculated from its two spatial derivatives, $g^x(x,y)$ and $g^y(x,y)$, using the identity: 
\begin{equation}
    \label{eqn:FDsoln}
    g(x,y) = \mathcal{F}^{-1}\left[\frac{\mathcal{F}\left(g^x(x,y)+ig^y(x,y)\right)}{ik_x-k_y}\right].
\end{equation}
Here, $\mathcal{F}$ denotes the two-dimensional Fourier transformation with respect to \textit{x} and \textit{y}, with $k_x$ and $k_y$ being the corresponding Fourier-space variables, respectively. This expression is unstable near the Fourier-space origin, $(k_x, k_y) = (0, 0)$,  hence the solution will diverge at low spatial frequencies unless sufficiently regularised. An effective diffusion coefficient can be calculated by equation~(\ref{eqn:FDsoln}), using the two calculated spatial derivatives $D_{\textrm{eff, Phase}}^x(\textbf{r};\Delta)$ and $D_{\textrm{eff, Phase}}^y(\textbf{r};\Delta)$. This solution will be stable at high spatial frequencies, that is, far from the Fourier-space origin, and the opposite is true for the effective diffusion coefficient $D_{\textrm{eff, Phase}}(\textbf{r};\Delta)$ that is extracted directly from the system of linear equations (equations~(\ref{eqn:FPPhaseCoupleSysa})\textendash(\ref{eqn:FPPhaseCoupleSysd})). These two solutions can be combined to consider only the stable spatial frequencies in each DF solution. In particular, a true effective diffusion coefficient, $D_{\textrm{eff, Phase}}^{\textrm{True}}(\textbf{r};\Delta)$, can be calculated by implementing the following Fourier-space weighted filtering (cf.~Ref.~\citenum{PaganinQPAM3}), where $\rho$ is a cut-off parameter:
\begin{equation}
\label{eqn:DFoptimal}
D_{\textrm{eff, Phase}}^{\textrm{True}}(\textbf{r};\Delta) = \mathcal{F}^{-1}\left[e^{-\rho(k_x^2+k_y^2)}\mathcal{F}\left(D_{\textrm{eff, Phase}}(\textbf{r};\Delta)\right) + \frac{1-e^{-\rho(k_x^2+k_y^2)}}{ik_x-k_y}\mathcal{F}\left(D_{\textrm{eff, Phase}}^x(\textbf{r};\Delta)+iD_{\textrm{eff, Phase}}^y(\textbf{r};\Delta)\right)\right].
\end{equation}
\\\\
This completes our description of the method for reconstructing a phase-object's true effective diffusion coefficient. The sample-induced phase-shift term, $\phi_{\textrm{ob}}(\textbf{r})$, can then be reconstructed by utilising equation~(\ref{eqn:FPPhaseSim}), via
\begin{equation}
\label{eqn:Phi}
\phi_{\textrm{ob}}(\textbf{r}) = \nabla_{\perp}^{-2}\left[\frac{k}{\Delta I_{R}(\textbf{r})}\left(I_R(\textbf{r})-I_{S}(\textbf{r})+\Delta\nabla_{\perp}^{2}\left[D_{\textrm{eff, Phase}}^{\textrm{True}}(\textbf{r};\Delta)I_{R}(\textbf{r})\right]\right)\right].
\end{equation}
In the above expression,
\begin{equation}
\label{eq:InverseLaplacianOperator}
\nabla_{\perp}^{-2} = -\mathcal{F}^{-1}\frac{1}{k_x^2+k_y^2}\mathcal{F}
\end{equation}
is the inverse Laplacian operator, derived using the two-dimensional Fourier derivative theorem\cite{paganin2006coherent}. The above PC signal extraction is more numerically stable than utilising the $\nabla_{\perp}^2\left[\frac{1}{k}\phi_{\textrm{ob}}(\textbf{r})-D_{\textrm{eff, Phase}}(\textbf{r};\Delta)\right]_{\textrm{Recon}}$ term reconstructed from QR decomposition\cite{press}, as that solution would inherently suffer from the numerical instabilities associated with $D_{\textrm{eff, Phase}}(\textbf{r};\Delta)$, $D_{\textrm{eff, Phase}}^x(\textbf{r};\Delta)$ and $D_{\textrm{eff, Phase}}^y(\textbf{r};\Delta)$, whereas equation~(\ref{eqn:Phi}) considers the stabilised $D_{\textrm{eff, Phase}}^{\textrm{True}}(\textbf{r};\Delta)$.
\\\\
Up until this point, X-ray attenuation by the sample has been neglected. We now extend our analysis to the case of a weakly-attenuating object and calculate its effective diffusion coefficient.  We do this based on a relationship obtained in Alloo \textit{et al.}\cite{alloo2022dark} (see equations~(24) and (18) therein) in which an attenuating-object's effective diffusion coefficient, $D_{\textrm{eff, Atten}}^{\textrm{True}}(\textbf{r};\Delta)$, can be calculated from the phase-object approximation, $D_{\textrm{eff, Phase}}^{\textrm{True}}(\textbf{r};\Delta)$, using
\begin{equation}
\label{DFattenphase}
D_{\textrm{eff, Atten}}^{\textrm{True}}(\textbf{r};\Delta) = \frac{D_{\textrm{eff, Phase}}^{\textrm{True}}(\textbf{r};\Delta)}{I_{\textrm{ob}}(\textbf{r})}.
\end{equation}
Above, $I_{\textrm{ob}}(\textbf{r})$ is the object's attenuation term describing the intensity at the exit surface of the sample, $z=0$, after the object has attenuated the incident X-ray beam of unit intensity. To calculate $I_{\textrm{ob}}(\textbf{r})$, we consider a single-material object such that the projection approximation\cite{paganin2006coherent} can be written as $\phi_{\textrm{ob}}(\textbf{r}) = -k \delta  t(\textbf{r})$ and $I_{\textrm{ob}}(\textbf{r}) = \textrm{exp}\left[-2k\beta t(\textbf{r})\right]$, where $t(\textbf{r})$ is the projected thickness of the object along the direction $z$ of the X-rays\cite{paganin2002simultaneous}.  Hence
\begin{equation}
\label{eqn:Iob}
I_{\textrm{ob}}(\textbf{r}) = \textrm{exp}\left[\frac{2\phi_{\textrm{ob}}(\textbf{r})}{\gamma}\right],
\end{equation}
where $\gamma = \delta/\beta$ for the single-material object. This attenuation term, obtained using the phase-shift term from equation~(\ref{eqn:Phi}), can then be used in equation~(\ref{DFattenphase}) to reconstruct $D_{\textrm{eff, Atten}}^{\textrm{True}}(\textbf{r};\Delta)$. Although in theory this attenuation extraction is restricted to single-material objects, it can be extended to multi-material objects by taking the difference in the refractive index components\cite{gureyev2002quantitative, beltran2010} for composite materials. Furthermore, in a tomographic context, it has been proven that this approximation does not affect the reconstructed attenuation coefficient, $\beta({\mathbf{r}}')$, far away from material interfaces\cite{gureyev2013accuracy}, and hence, this restriction would only be adverse in a sample that has several composite materials with significantly differing attenuation and refraction properties.

\section*{Stabilising the SB-PCXI multimodal inverse-problem}

The inverse-problem\cite{Sabatier2000} of reconstructing the effective diffusion coefficient requires appropriate numerical regularisation. Here, we apply the common approach\cite{fathi2018deep,murli1999wiener} of a Tikhonov regularisation\cite{tikhonov1977solutions}, which sufficiently stabilises the signal reconstruction. In its simplest form, a Tikhonov regularisation of the quotient of two functions $A$ and $B$ can be employed, using
\begin{equation}
\label{eqn:TR}
    \frac{A}{B} \rightarrow \frac{AB}{B^2+\alpha},
\end{equation}
where $\alpha \ge 0$ is a regularisation parameter whose magnitude is sufficiently small compared to $B^2$.
\\\\
Within this paper, the effective diffusion coefficient is reconstructed by solving a full-rank system of four linear equations, equations~(\ref{eqn:FPPhaseCoupleSysa})\textendash(\ref{eqn:FPPhaseCoupleSysd}). The numerical stability of this solution can be improved by utilising SB-PCXI data from more mask positions. Namely, for \textit{N} mask positions, we can generate a system of \textit{N} linear equations following the form of equation~(\ref{eqn:FPPhaseSimEx}). The system generated can then be solved in a least-squares sense using pixel-wise QR decomposition\cite{press}. Tikhonov's regularisation method can also be applied to an ill-posed least-squares problem (QR factorisation), as described by Zhu\cite{zhu1993qr}. Namely, rather than using QR decomposition to solve the linear system $A\bar{x} = \bar{b}$ for the least-squares solution $\tilde{x}$, QR decomposition can instead be performed on the system $\begin{pmatrix}A;\alpha I\end{pmatrix}\bar{x}=\begin{pmatrix}\bar{b};0\end{pmatrix},$ where $M \times N$ is the coefficient matrix $A$, ``;'' denotes a new row, $I$ is the $N \times N$ identity matrix, $\alpha$ is the chosen regularisation parameter, and the right-hand-side vector $\bar{b}$ is filled with zeroes to reach the size of $(M + N) \times 1$. The described Tikhonov-regularised QR decomposition can be used to solve the overdetermined system of linear equations for the four unknown variables, $\nabla_{\perp}^2\left[\frac{1}{k}\phi_{\textrm{ob}}(\textbf{r})-D_{\textrm{eff, Phase}}(\textbf{r};\Delta)\right]$, $D_{\textrm{eff, Phase}}(\textbf{r};\Delta)$, $D_{\textrm{eff, Phase}}^x(\textbf{r};\Delta)$, and $D_{\textrm{eff, Phase}}^y(\textbf{r};\Delta)$. For reasons described in the preceding text, equation~(\ref{eqn:DFoptimal}) is numerically stable and therefore does not need to be regularised. The subsequent phase extraction, that is equation~(\ref{eqn:Phi}), is ill-posed close to the Fourier-space origin $(k_x, k_y) = (0, 0)$ and hence an appropriate Tikhonov regularisation should be applied following equation~(\ref{eqn:TR}). For severely ill-posed cases, the phase extraction can  be further stabilised by utilising instances of equation~(\ref{eqn:FPPhaseSim}), substituting in the true effective diffusion coefficient using the method described above, and performing Tikhonov-regularised QR decomposition to solve for $\nabla_{\perp}^{-2}\phi_{\textrm{ob}}(\textbf{r})$ before applying the inverse Laplacian operator to reconstruct $\phi_{\textrm{ob}}(\textbf{r})$.
\\\\
For the Tikhonov regularisation to operate successfully, the regularisation parameter, $\alpha$, needs to be selected appropriately for the given data. If $\alpha$ is chosen to be too large, the computed solution will be over-smoothed and will therefore lack fine detail. In the case of image reconstruction, this means the computed solution will have poor spatial resolution, although a high SNR. Conversely, if $\alpha$ is chosen to be too small, the computed solution will be severely contaminated with errors resulting from numerical instabilities. Evidently, optimising the regularisation parameter is critical to successfully extract multimodal signals. There are algorithms in the current research literature that optimise the Tikhonov regularisation parameter for a given ill-posed problem, see Park \textit{et al.}\cite{park2018parameter} and references therein. However, in image reconstruction, image quality metrics can be used to determine the optimal regularisation parameter. In the present work, we used four metrics:
\begin{enumerate}
  \item Naturalness Image Quality Evaluator (NIQE)\cite{mittal2012making}: The NIQE is a blind image quality assessment based on an image's measurable deviations from statistical regularities observed in natural images. A lower NIQE reflects an image with a higher perceived image quality. 
  
  \item Azimuthally Averaged Power-Spectrum\cite{nill1976scene}: The two-dimensional power-spectrum of an image can be calculated by taking the absolute square of the Fourier-transformed image. This can then be azimuthally-averaged, with the centre at the Fourier-space origin $(k_x, k_y) = (0, 0)$, to calculate a one-dimensional power-spectrum that shows the contribution of all spatial frequencies in an image. The noise in the image is reflected by the so-called ``noise-floor'', which typically makes up the majority of the signal at high spatial frequencies, and the spatial resolution can be gauged by the ``knee'' of the power-spectrum, namely the frequency at which noise becomes a significant contribution to the signal. 
  \item SNR: The SNR measures the magnitude of a signal relative to background noise. It is used to quantify signal quality in an image, and is defined as 
    \begin{equation}
        SNR = I_{\textrm{avg}}/\sigma,
    \end{equation}
    where $I_{\textrm{avg}}$ is the signal strength, which can be measured as the average pixel value within a region of approximately uniform signal, and $\sigma$ is the noise as measured by the standard deviation of pixel values. Note that if the noise characteristics are the same inside and outside an object, then it may be easiest to measure $\sigma$ in the region outside the object to avoid variations in the signal that come with a complex object.
  \item Human visual perception: Although subjective, meaningful image quality measurements can be made by a human observer's visual assessment of an image. 
\end{enumerate}

\section*{Applying the theoretical approach to synchrotron SB-PCXI data}

To test the proposed MIST approach, we extracted multimodal signals from two samples, a wattle flower (WF) and red currant (RC). These two samples had different X-ray attenuation coefficients, $\beta(\textbf{r}')$, and thicknesses, and hence attenuate the X-ray beam differently. In particular, the WF was weakly-attenuating (almost a pure phase-object) and the RC was non-negligibly-attenuating.

\subsection*{Experimental procedures}

SB-PCXI data of the WF were collected in experimental hutch 3B of the Imaging and Medical Beamline (IMBL) at the Australian Synchrotron, similar to the setup shown in Fig.~\ref{fig:setup}. The entrance window to hutch 3B was located 135 m from the source. A virtually monochromatic 25 keV X-ray beam was used for imaging. The Ruby detector\cite{stevenson2017quantitative}, which has a single pco.edge sensor and lens-coupled scintillator, was positioned $\Delta = 2$ m downstream from the sample. The pixel array was 2560 $\times$ 2160 and a 105 mm macro lens was used to achieve an effective pixel size of 9.9 $\mu$m. The speckle-generating mask was located around 60 cm upstream of the sample, and a combination of sandpapers of grit P40, P80 and P120 were simultaneously used. A combination of sandpapers was used in this case such that the generated speckle pattern had a range of high-visibility feature sizes. The speckle-generating sandpaper was placed on a translation motor stage such that it could be brought in and out of the X-ray beam path and also translated transversely to allow the beam to pass through different parts of the mask. The effective speckle-size was 136.4 $\mu$m, as measured by the average full-width at half-maximum of the autocorrelation function\cite{goodman2020speckle}  in the horizontal, \textit{x}, and vertical, \textit{y}, directions of the reference-speckle field.
\\\\The RC sample was imaged using SB-PCXI at the European Synchrotron Radiation Facility (ESRF) beamline BM05; these data were obtained and originally published by Berujon and Ziegler\cite{berujon2016x}. The setup was similar to that shown in Fig.~\ref{fig:setup}, with an X-ray energy of 17 keV and spectral bandwidth of $\Delta E/E \approx 10^{-4}$, which was produced using a double crystal Si(111) monochromator located 27 m from the X-ray source. The RC was placed on a stage 55 m from the source. The detector system consisted of a Fast Read-Out Low-Noise (FReLoN) e2V camera coupled to an optical imaging thin scintillator\cite{labiche1996frelon,douissard2012versatile}. This detector was placed $\Delta = 1$ m downstream from the sample and the effective pixel size of the optical system was 5.8 $\mu$m. The speckle-generating sandpaper with grit size P800 was placed 0.5 m upstream from the sample and had an effective speckle-size of 20.4 $\mu$m.
\\\\
Image acquisitions for all data were collected using a similar procedure for both samples: dark-current (no X-ray beam) and flat-field (sample and mask not in the beam) exposures were collected before and after the scan, and reference-speckle images with only the speckle mask in the beam were collected before and after the scan. The speckle mask was transversely shifted in the $x$-direction perpendicular to the optical axis (see Fig.~\ref{fig:setup}) to acquire multiple unique sets of SB-PCXI data. Seven and fifteen sets of SB-PCXI data were collected for the RC and WF, respectively. The collected SB-PCXI data were then processed using a Python3 script to implement our  multimodal signal extraction algorithm. An open-access repository for this code is on GitHub\cite{GITHUB}.

\subsection*{Multimodal signal extraction}

Multimodal signals were extracted for the WF and RC using our new generalised MIST approach. The entirety of the available SB-PCXI data for each sample, that is the maximum number of masks, were used to reconstruct the multimodal signals. Although fewer could be used, this work focuses solely on the new theoretical development; a quantitative analysis of the influence of the number of SB-PCXI data sets is given in Pavlov \textit{et al.}\cite{pavlov2020x} (see, in particular, Fig.~2 in Ref. \citenum{pavlov2020x}). The established system of linear equations for each sample was solved using the Tikhonov-regularised QR decomposition described above. It was found that the standard deviation of the coefficient matrix divided by $10^{4}$ provided the optimal Tikhonov regularisation parameter for the case of each sample. The phase-object approximation of the sample's effective diffusion coefficient, $D_{\textrm{eff, Phase}}(\textbf{r};\Delta)$, and its spatial derivatives were calculated using this method, from which the true phase-object approximation of the DF signal, $D_{\textrm{eff, Phase}}^{\textrm{True}}(\textbf{r};\Delta)$, was computed. Next, the sample's phase-shifts, $\phi_{\textrm{ob}}(\textbf{r})$, attenuation term, $I_{\textrm{ob}}(\textbf{r})$, and true attenuating-object effective diffusion coefficient, $D_{\textrm{eff, Atten}}^{\textrm{True}}(\textbf{r};\Delta)$, were calculated. It should be noted that the calculation of $\phi_{\textrm{ob}}(\textbf{r})$ had to be regularised as equation~(\ref{eq:InverseLaplacianOperator}) is unstable at the Fourier-space origin;  it was regularised using equation~(\ref{eqn:TR}) with $\alpha = 0.0001$, which was suitable in all cases. The variable $\gamma = \delta/\beta$ was required to calculate the object's attenuation term, as in equation~(\ref{eqn:Iob}). For the WF, the generic elemental composition of a plant stem was used within the TS imaging calculator\cite{TSIMAGING} to determine its complex refractive index at 25 keV, producing $\gamma_{\,\textrm{WF, 25~keV}} = 1403$. For the RC sample, $\gamma$ was taken to be that of water at 17 keV, that is, $\gamma_{\,\textrm{RC, 17~keV}} = \delta_{\,\textrm{water, 17~keV}}/\beta_{\,\textrm{water, 17~keV}} = 1146$.
\\\\\
\noindent Multimodal signals were also calculated using our previously-published MIST approaches, that is, equation~(6) in Pavlov \textit{et al.}\cite{pavlov2020x} and equations~(16)\textendash(18) in Alloo \textit{et al.}\cite{alloo2022dark}, to provide a point of comparison for the new approach. As described earlier, the approach published by Pavlov \textit{et al.}\cite{pavlov2020x} neglects X-ray attenuation, but Alloo \textit{et al.}\cite{alloo2022dark} considers it. In both of these approaches, the effective diffusion coefficient is assumed to be spatially slowly varying. The multimodal signal extraction described by Alloo \textit{et al.}\cite{alloo2022dark} involved numerical stabilisation via a Tikhonov-regularised ``Weighted Determinant'' approach. Within this work, the multimodal signals extracted using Pavlov \textit{et al.}'s\cite{pavlov2020x} algorithm were stabilised in an identical way to that described in Alloo \textit{et al.}\cite{alloo2022dark}, that is a Tikhonov-regularised ``Weighted Determinant'' approach. The optimal regularisation parameter, in both instances, was equal to the mean of the denominator in equation~(23) of Ref.~\citenum{alloo2022dark}, divided by 100, for both samples. Here we do not perform a comparison with our directional DF approach\cite{pavlov2021directional}.

\subsection*{Weakly-attenuating sample: Wattle flower (WF)}

\begin{figure}[h]
    \centering
    \includegraphics[width=0.65\linewidth]{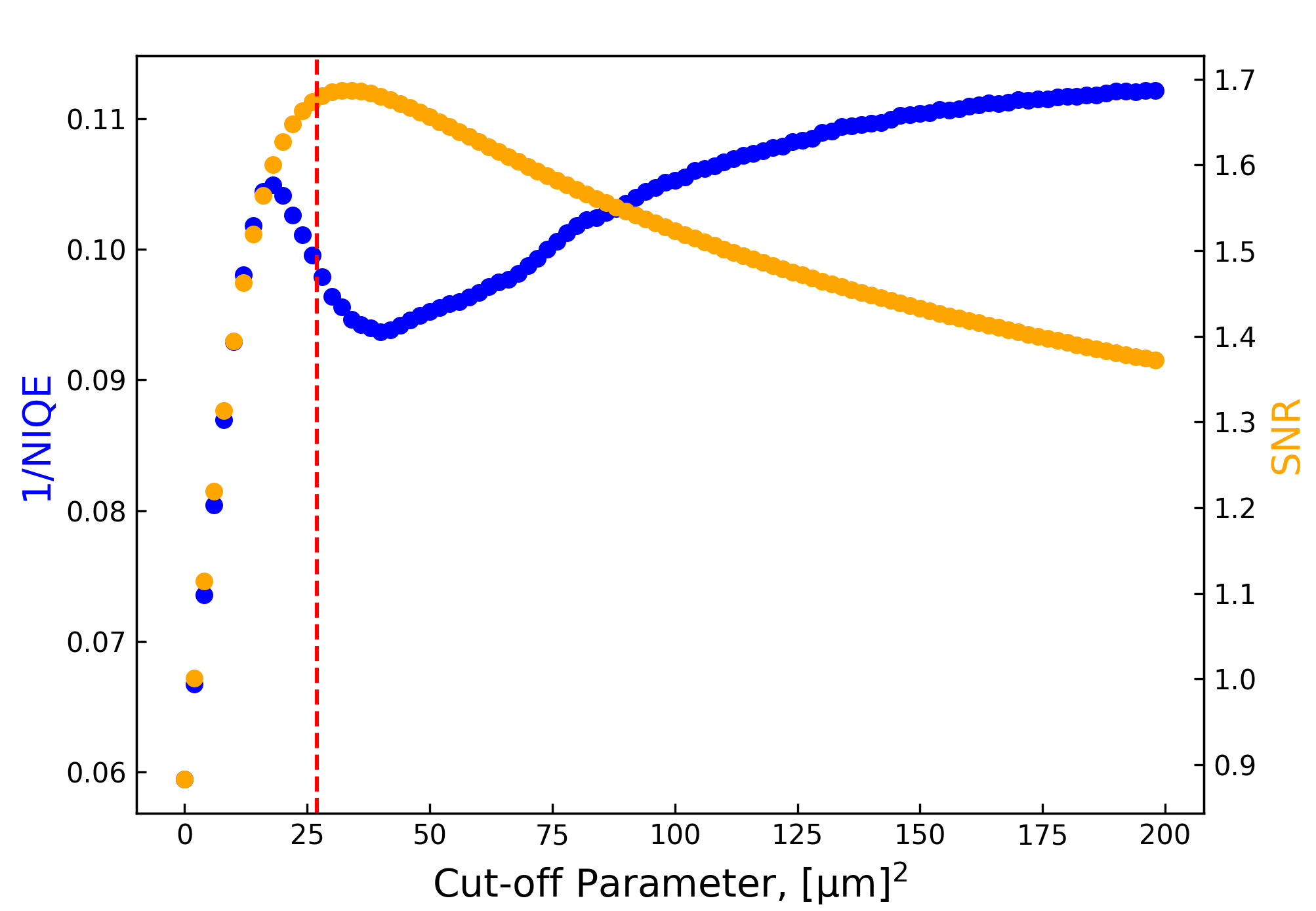}
    \caption{\textit{Influence of cut-off parameter, $\rho$, on the image quality of the wattle flower's reconstructed phase-object approximation of the effective diffusion coefficient, $D_{\textrm{eff, Phase}}^{\textrm{True}}(\textbf{r};\Delta)$. Image quality is measured by the (blue) reciprocal of the Naturalness Image Quality Evaluator (NIQE)\cite{mittal2012making} and (orange) signal-to-noise ratio (SNR). The dashed red vertical line denotes the optimal cut-off parameter that appropriately considers both metrics and the verdict of a human observer. 
    }}
    \label{fig:varyCutt}
\end{figure}

We begin by investigating the WF sample as this sample conforms most closely to the underlying assumptions of the derived theoretical approach, as it is a weakly-attenuating object. Therefore, the multimodal signals (DF and PC) should be superior using the present approach compared to those extracted using our previous MIST approaches\cite{pavlov2020x,alloo2022dark}. Using the methodology described above, an over-determined system of fifteen linear equations was solved using Tikhonov-regularised QR decomposition, from which the true phase-object DF approximation, $D_{\textrm{eff, Phase}}^{\textrm{True}}(\textbf{r};\Delta)$, attenuation term, $I_{\textrm{ob}}(\textbf{r})$, and true attenuating-object DF signals, $D_{\textrm{eff, Atten}}^{\textrm{True}}(\textbf{r};\Delta)$, were calculated. The cut-off parameter, $\rho$, was determined by investigating the SNR and NIQE of the reconstructed $D_{\textrm{eff, Phase}}^{\textrm{True}}(\textbf{r};\Delta)$ for various cut-off parameter values. Figure~\ref{fig:varyCutt} shows how the value of the cut-off parameter influences the image quality of the reconstructed DF, as measured by the NIQE and SNR of the entire DF reconstruction. As an image with a better-perceived image quality has a lower NIQE,  Fig.~\ref{fig:varyCutt} is shown as the reciprocal of the NIQE such that it follows the same ``bigger means better'' convention as the SNR shown on the secondary axis. The optimal cut-off parameter is 34 $\mu\textrm{m}{^2}$ and 198 $\mu\textrm{m}{^2}$ as measured by the SNR and reciprocal NIQE, respectively. As we have described, the NIQE is a blind image quality metric that measures the so-called perceived image quality. Within this work, the NIQE was calculated using an untrained NIQE function, called {\em niqe(A)}, in MATLAB. This untrained NIQE model may mistake artefacts for signal. For example, enhanced edges due to residual PC effects, or Fresnel fringes, are considered qualitatively better by the NIQE as they have an increased sharpness and contrast. Such features are present in the reconstructed DF signal when the cut-off parameter is too large, and hence, explains why the NIQEs for these images indicate higher perceived image quality. It follows that the optimal cut-off parameter should be selected by appropriately considering the SNR and NIQE simultaneously. We selected the optimal cut-off parameter to be $\rho = $27 $\mu\textrm{m}{^2}$ as this balances the local reciprocal NIQE maximum (at approximately $\rho = $18 $\mu\textrm{m}{^2}$), global SNR maximum, and also the human observer's verdict. 
\begin{figure}[h]
    \centering
    \includegraphics[width=1.0\linewidth]{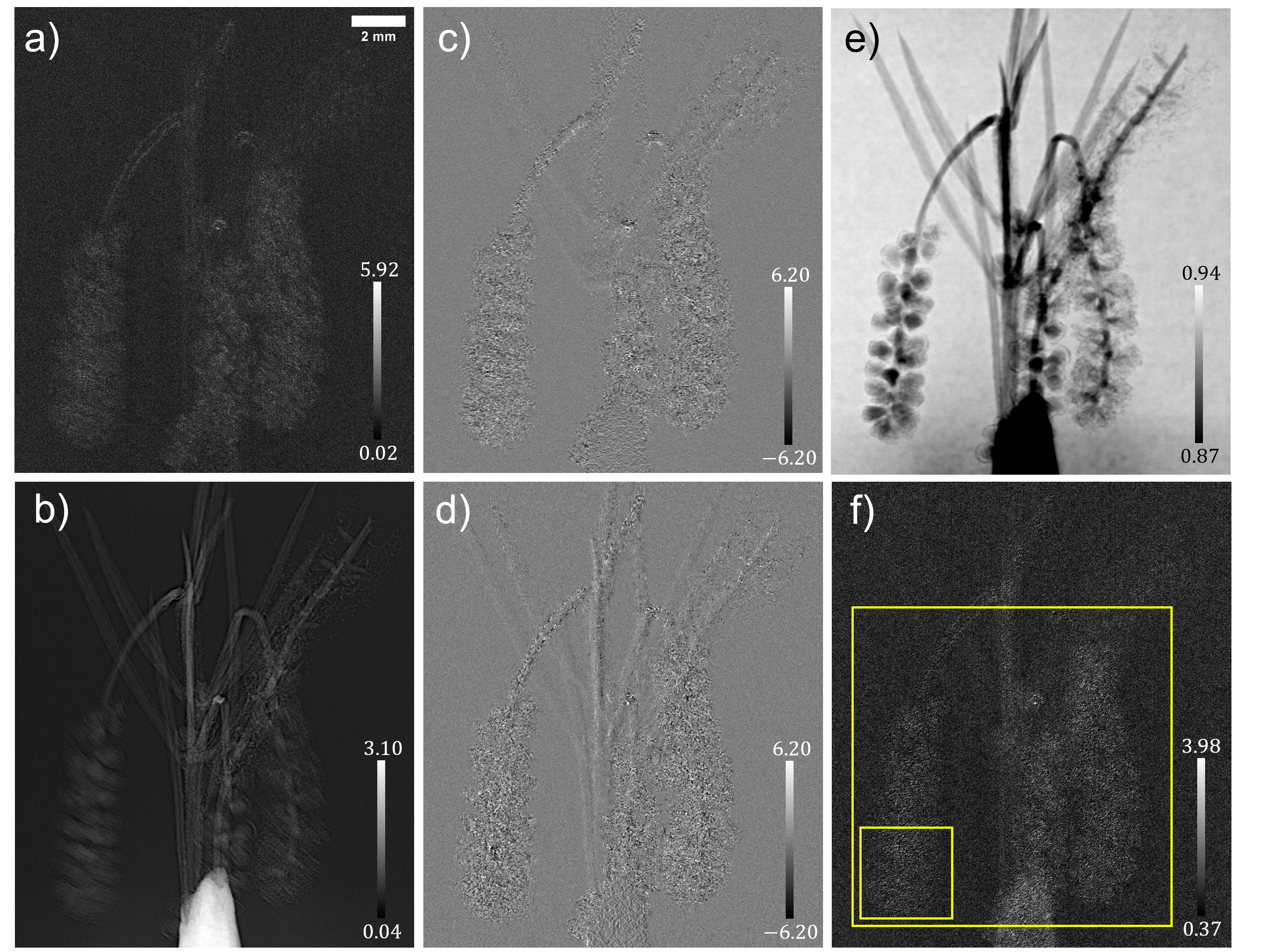}
    \caption{\textit{
    Solutions of the system of linear equations, equations~(\ref{eqn:FPPhaseCoupleSysa})\textendash(\ref{eqn:FPPhaseCoupleSysd}), and the wattle flower's reconstructed multimodal signals; (a) is the effective diffusion coefficient, $D_{\textrm{eff, Phase}}(\textbf{r};\Delta)$, (b) is $\nabla_{\perp}^2\left[\frac{1}{k}\phi_{\textrm{ob}}(\textbf{r})-D_{\textrm{eff, Phase}}(\textbf{r};\Delta)\right]_{\textrm{Recon}}$, (c) and (d) are the two spatial derivatives of the effective diffusion coefficient, $D_{\textrm{eff, Phase}}^y(\textbf{r};\Delta)$ and $D_{\textrm{eff, Phase}}^x(\textbf{r};\Delta)$, respectively, (e) is the reconstructed attenuation term, $I_{\textrm{ob}}(\textbf{r})$, and (f) is the wattle flower's optimally filtered true effective diffusion coefficient, $D_{\textrm{eff, Atten}}^{\textrm{True}}(\textbf{r};\Delta)$. The greyscale bars in subfigures (a) and (f) are $\times10^{-5}$ $\mu$m, (b) is $\times10^{-7}$ $\mu$m$^{-1}$, and (c) and (d) are $\times10^{-6}$. 
    }}
    \label{fig:WFSolutions}
\end{figure}
\begin{figure}[h]
    \centering
    \includegraphics[width=1.0\linewidth]{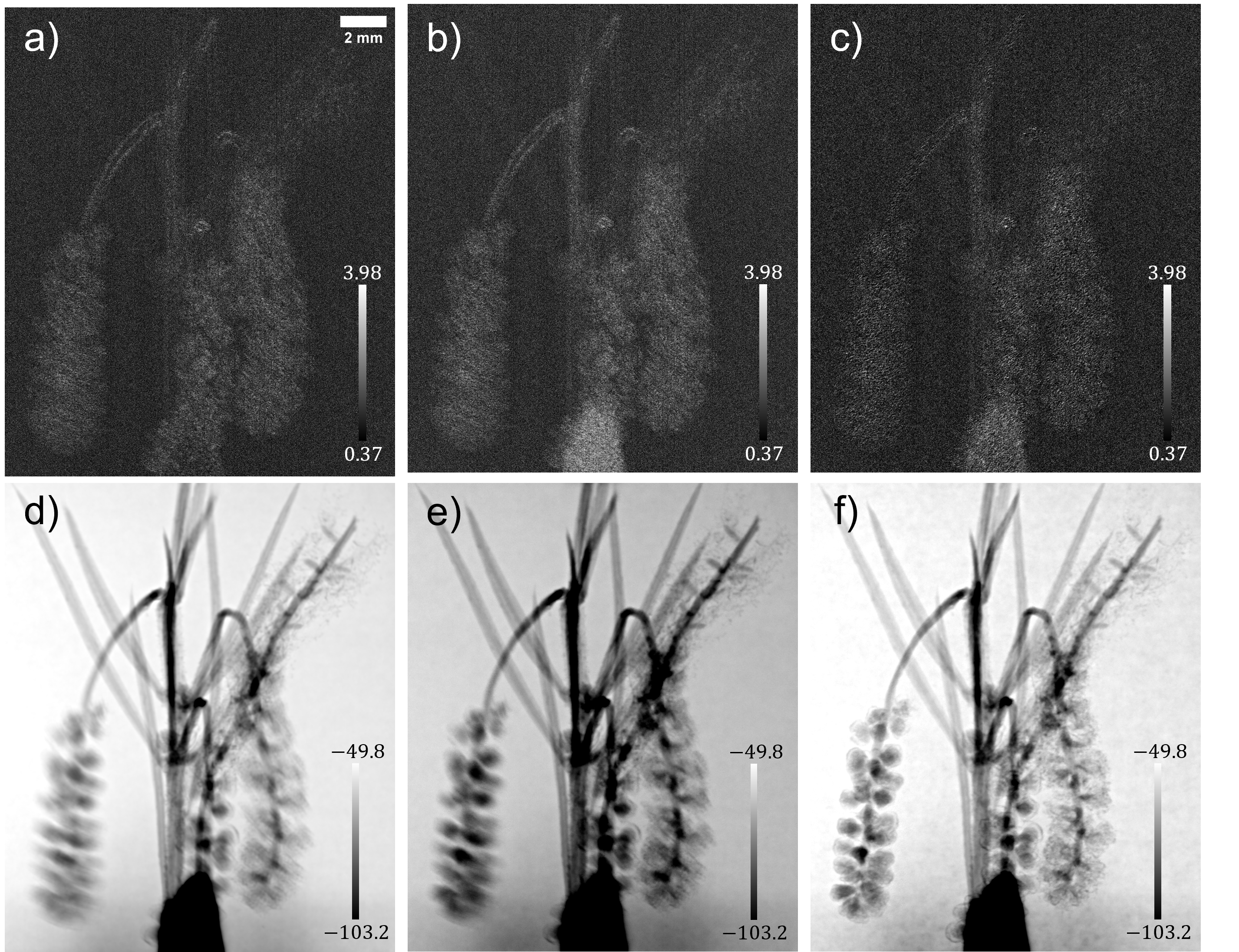}
    \caption{\textit{Comparison of  wattle-flower multimodal signals extracted using the MIST approaches for rotationally-isotropic diffuse scatter: (a)\textendash(c) are reconstructed effective diffusion coefficients, and (d)\textendash(f) are phase-shifts (as the wattle flower is a pure phase-object). (a) and (d) use Pavlov \textit{et al.}'s\cite{pavlov2020x} approach, which neglects X-ray attenuation, and (b) and (e) use Alloo \textit{et al.}'s\cite{alloo2022dark} approach, which considers it; both approaches assume the effective diffusion coefficient to be slowly varying. (c) and (f) are calculated using the present approach, which considers weak X-ray attenuation and does not approximate the diffusion coefficient as slowly varying. The greyscale bars in (a)\textendash(c) are $\times10^{-5}$ $\mu$m and (d)\textendash(f) are $\times10^{0}$ rads.}}
    \label{fig:WFSigs}
\end{figure}
\medskip
\newline Figure~\ref{fig:WFSolutions} shows the computed solutions, $\nabla_{\perp}^2\left[\frac{1}{k}\phi_{\textrm{ob}}(\textbf{r})-D_{\textrm{eff, Phase}}(\textbf{r};\Delta)\right]$, $D_{\textrm{eff, Phase}}(\textbf{r};\Delta)$, $D_{\textrm{eff, Phase}}^x(\textbf{r};\Delta)$, and $D_{\textrm{eff, Phase}}^y(\textbf{r};\Delta)$, and the multimodal signals, $I_{\textrm{ob}}(\textbf{r})$ and $D_{\textrm{eff, Atten}}^{\textrm{True}}(\textbf{r};\Delta)$, for the WF sample, where the window and level of each image were set to optimise the respective greyscale range. Although this sample has a weak DF signal, Figs.~\ref{fig:WFSolutions}b\textendash\ref{fig:WFSolutions}d reveal the first- and second-order derivatives of the diffuse scattering signal that had been neglected in prior MIST approaches. It is obvious that the gradient term is stronger at material interfaces on a global and local scale, that is the WF leaf edges and the filaments that make up each flower, respectively. This follows our initial prediction that the assumption of the DF signal being slowly varying at material interfaces was insufficient; this is furthermore supported in the case of the RC sample which is discussed later. Here, we emphasise that the WF's reconstructed $I_{\textrm{ob}}(\textbf{r})$ is close to unity, confirming that it is a weakly-attenuating object.
\\\\
Figure~\ref{fig:WFSigs} compares the present MIST approach to those in the current MIST research literature\cite{pavlov2020x,alloo2022dark}; the reconstructed DF signals are shown in Figs.~\ref{fig:WFSigs}a\textendash\ref{fig:WFSigs}c and the phase-shifts in Figs.~\ref{fig:WFSigs}d\textendash\ref{fig:WFSigs}f. Figures~\ref{fig:WFSigs}a and \ref{fig:WFSigs}b are qualitatively equivalent with regard to the resolvability of the WF's leaves and filaments. Quantitatively the reconstructed DF signal in  Fig.~\ref{fig:WFSigs}b is larger than that in  Fig.~\ref{fig:WFSigs}a, owing to the former being a phase-object approximation and the latter considering X-ray attenuation, and hence there is a higher reconstructed DF signal in regions that attenuated the X-ray beam more; this agrees with what was found in Alloo \textit{et al.}~\cite{alloo2022dark}. The DF signal computed using the present approach, Fig.~\ref{fig:WFSigs}c, initially appears to have a weaker reconstructed signal within the WF's leaves and filaments. However, after inspecting the reconstructed phase-shifts using all MIST approaches, Figs.~\ref{fig:WFSigs}d\textendash\ref{fig:WFSigs}f, it becomes apparent the DF signals reconstructed using the alternative approaches have predominant blurring at the WF's flowers. Moreover, the WF's DF signal from the filaments that make up each flower cannot be well-described by assuming the DF is slowly varying, as assumed in Pavlov \textit{et al.}\cite{pavlov2020x} and Alloo \textit{et al.}\cite{alloo2022dark}. For this sample, the previous MIST formalisms are unable to detect the rapidly varying features inside the WF filaments and its edges. This is apparent in the reconstructions of the phase-shifts and effective diffusion signal. This weakly-attenuating sample demonstrates that our new approach gives reconstructed multimodal signals that are qualitatively superior compared to the previously-published MIST approaches. Note that a quantitative comparison of the reconstructed signals' image quality is provided in a following section.

\begin{figure}[h]
    \centering
    \includegraphics[width=\linewidth]{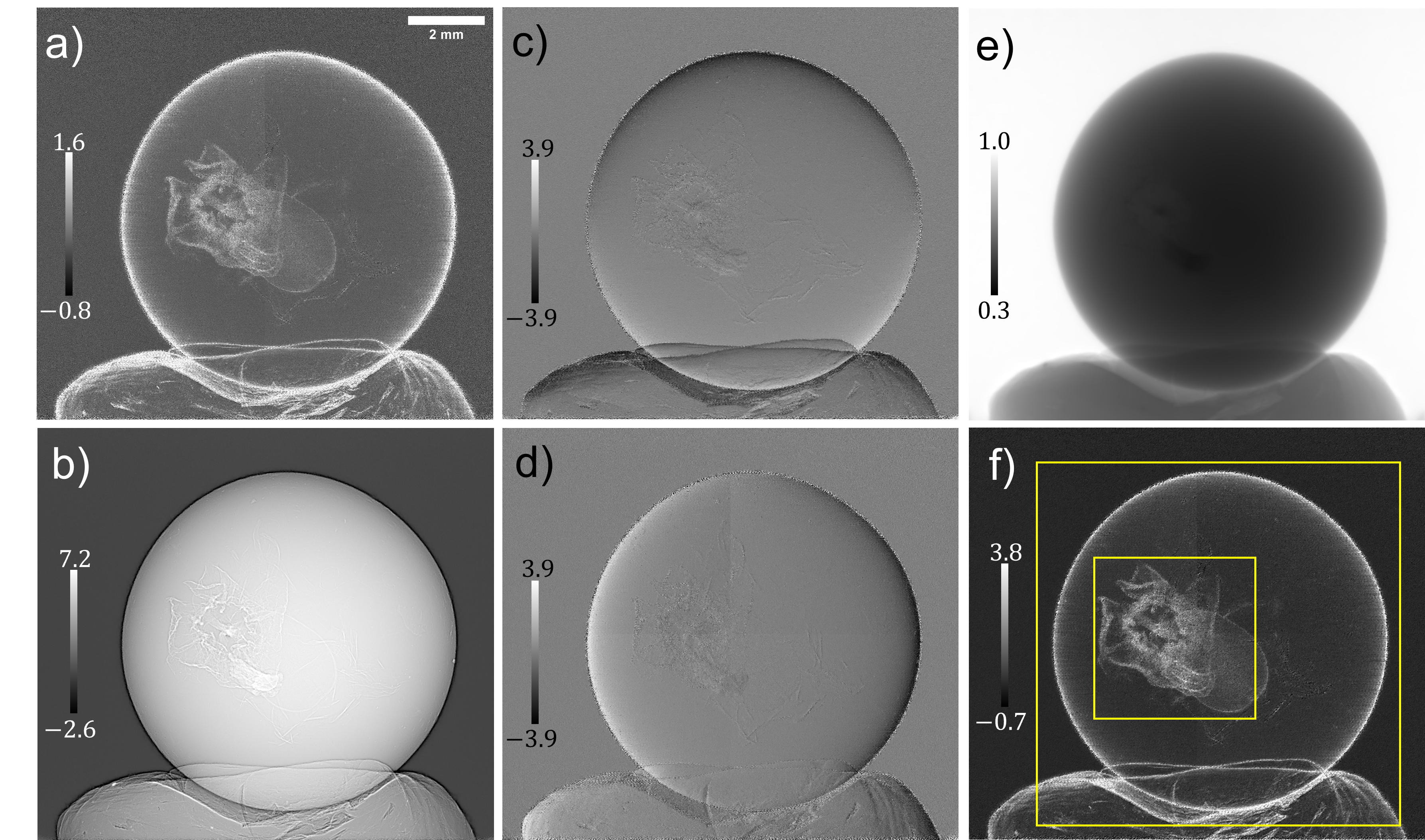}
    \caption{\textit{Solutions of the system of linear equations, equations~(\ref{eqn:FPPhaseCoupleSysa})\textendash(\ref{eqn:FPPhaseCoupleSysd}), and the red currants's reconstructed multimodal signals; (a) is the effective diffusion coefficient, $D_{\textrm{eff, Phase}}(\textbf{r};\Delta)$, (b) is $\nabla_{\perp}^2\left[\frac{1}{k}\phi_{\textrm{ob}}(\textbf{r})-D_{\textrm{eff, Phase}}(\textbf{r};\Delta)\right]_{\textrm{Recon}}$, (c) and (d) are the two spatial derivatives of the effective diffusion coefficient, $D_{\textrm{eff, Phase}}^y(\textbf{r};\Delta)$ and $D_{\textrm{eff, Phase}}^x(\textbf{r};\Delta)$, respectively, (e) is the reconstructed attenuation term, $I_{\textrm{ob}}(\textbf{r})$, and (f) is the red currants's optimally filtered true effective diffusion coefficient, $D_{\textrm{eff, Atten}}^{\textrm{True}}(\textbf{r};\Delta)$.
    The greyscale bars in subfigures (a) and (f) are $\times10^{-5}$ $\mu$m, (b) is $\times10^{-7}$ $\mu$m$^{-1}$, and (c) and (d) are $\times10^{-6}$. 
    }}
    \label{fig:RC_soln}
\end{figure}

\subsection*{Attenuating sample: Red currant (RC)}

\begin{figure}[h]
    \centering
    \includegraphics[width=\linewidth]{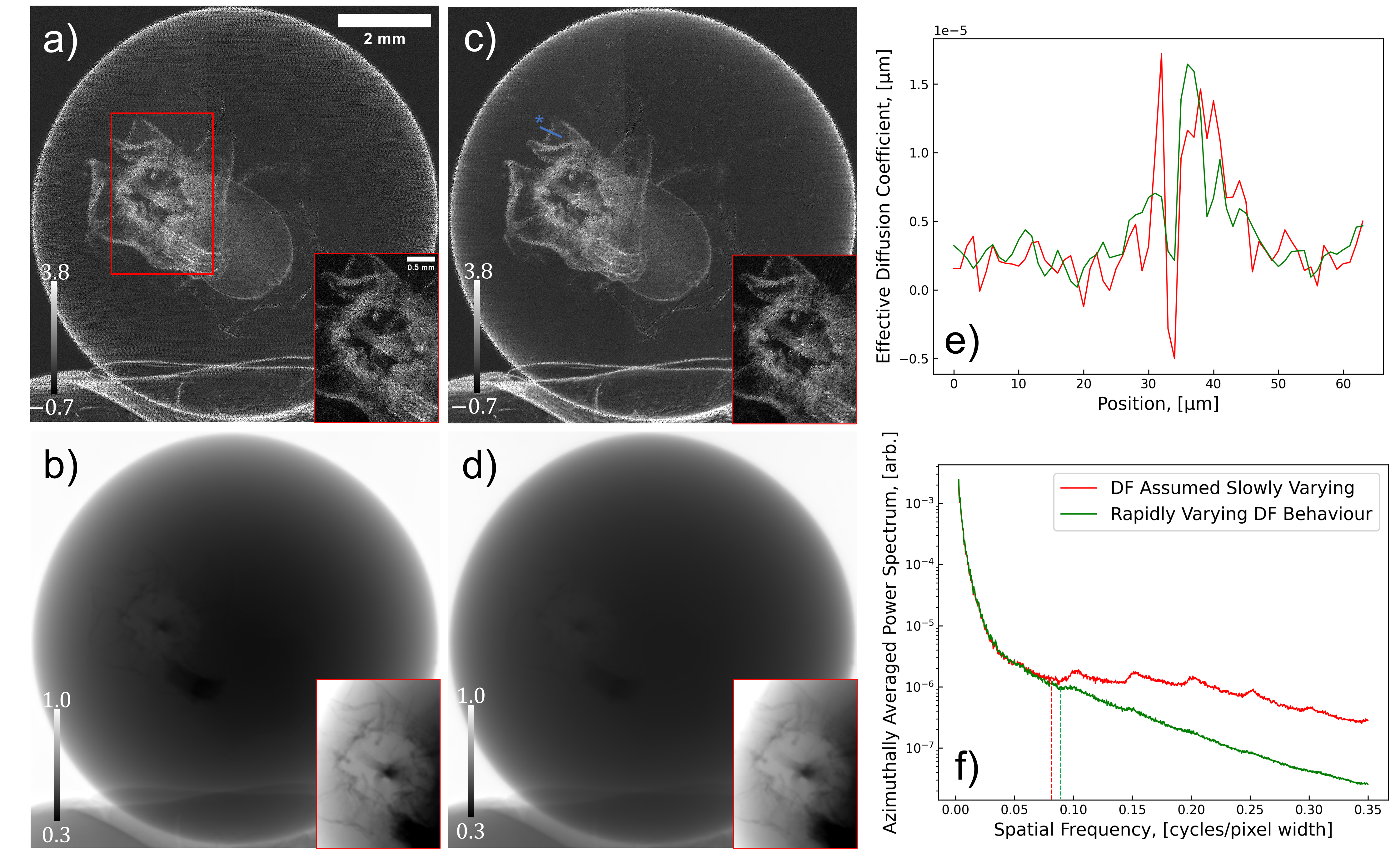}
    \caption{\textit{Comparison of the red currant's multimodal signals extracted using the approach published in Alloo \textit{et al.}\cite{alloo2022dark}, (a)\textendash(b), and that presented within this study, (c)\textendash (d). Figures  (a)\textendash(b) assume the red currant's effective diffusion coefficient is slowly varying and Figs. (c)\textendash(d) have no assumptions regarding the effective diffusion coefficient. The effective diffusion coefficient reconstructions are shown in (a) and (c), and the RC's attenuation term is shown in (b) and (d). The blue-asterisk line profile in (c) is shown in (e), and (f) is the azimuthally averaged power spectra. The vertical dashed lines in (f) denote the approximate knee of the respective power spectra. The greyscale bars in (a) and (c) are $\times10^{-5}$ $\mu$m.
    }}
    \label{fig:RC_crop}
\end{figure}

We now turn to the non-negligibly-attenuating RC sample, which tests the breadth of applicability of the proposed approach. Using an identical methodology to that described for the WF, the over-determined system of linear equations for seven sets of SB-PCXI intensity data was solved using Tikhonov-regularised QR decomposition, from which the RC's true phase-object DF approximation, $D_{\textrm{eff, Phase}}^{\textrm{True}}(\textbf{r};\Delta)$, attenuation term, $I_{\textrm{ob}}(\textbf{r})$, and true attenuating-object DF, $D_{\textrm{eff, Atten}}^{\textrm{True}}(\textbf{r};\Delta)$, signals were calculated. The cut-off parameter was chosen to be $\rho =$ 21 $\mu\textrm{m}{^2}$ by appropriately optimising the NIQE and SNR, as described earlier. 
\\\\
Figure~\ref{fig:RC_soln} shows all of the relevant solutions and multimodal signals for the RC sample using the above-described method; Figs.~\ref{fig:RC_soln}a\textendash\ref{fig:RC_soln}d show the reconstructed solutions of the system of linear equations, and Figs.~\ref{fig:RC_soln}e\ and \ref{fig:RC_soln}f display the RC's attenuation term and true attenuating-object approximation of its effective diffusion coefficient, respectively. Similar to the case of the WF, the gradient term is stronger at the periphery of the entire RC sample and at the internal fibres. Figure~\ref{fig:RC_soln}e shows the RC’s attenuation term, where the microstructure is indistinguishable due to its attenuation-based contrast/signal being small. Such microstructure, which is unresolved in the attenuation term, induces measurable SAXS, and hence, can be readily resolved in the reconstructed DF signal shown in Fig.~\ref{fig:RC_soln}f. It is important to make note of the RC's reconstructed $I_{\textrm{ob}}(\textbf{r})$, particularly how far from unity it is. This indicates that the attenuating-object approximation for the effective diffusion coefficient should be used to reconstruct a true DF signal. 
\\\\
Figure~\ref{fig:RC_crop} gives a direct comparison of the RC's DF and attenuation reconstructions, using (a) the most recently published rotationally-isotropic-scatter ``DF Assumed Slowly Varying'' MIST approach\cite{alloo2022dark} (Figs.~\ref{fig:RC_crop}a and \ref{fig:RC_crop}b), and (b) the current ``Rapidly Varying DF Behaviour'' MIST approach (Figs.~\ref{fig:RC_crop}c and \ref{fig:RC_crop}d). Here, we only compare the present approach to the rotationally-isotropic MIST approach\cite{alloo2022dark} that considers X-ray attenuation, as that published by Pavlov \textit{et al.}\cite{pavlov2020x} would erroneously reconstruct the RC's DF signal since the sample significantly attenuates the X-ray beam. Figures~\ref{fig:RC_crop}a and \ref{fig:RC_crop}c show a local reconstruction of the RC's $D_{\textrm{eff, Atten}}^{\textrm{True}}(\textbf{r};\Delta)$ and Figs.~\ref{fig:RC_crop}b and \ref{fig:RC_crop}d are the $I_{\textrm{ob}}(\textbf{r})$ reconstructions; magnified regions of each extracted signal are also shown. By comparing the reconstructed DF signals, it is evident that the new approach increases spatial resolution, decreases noise, and also provides a greater subject-contrast between unresolved microstructure. Moreover, the small fibrous network surrounding the RC pip, otherwise known as the pericarp, is more resolvable in Fig.~\ref{fig:RC_crop}c than Fig.~\ref{fig:RC_crop}a. A line profile across a single RC fibre, indicated by the blue asterisk in Fig.~\ref{fig:RC_crop}c, is shown in Fig.~\ref{fig:RC_crop}e. This fine feature is resolvable in the DF image reconstructed using the current approach (denoted by the green trace), but not when the DF is assumed to be slowly varying (denoted by the red trace). When the DF is assumed to be slowly varying, there are supposedly Fresnel fringes, or residual PC, at the boundaries of the traced pericarp fibre in the recovered DF signal; this is the high-low intensity region at approximately 30 $\mu m$ and 35 $\mu m$ in Fig.~\ref{fig:RC_crop}e. When this assumption is relaxed, the apparent DF signal induced by the strong phase effects is reconstructed appropriately, such that the small fibre is resolvable. This line profile indicates the evident spatial resolution difference between the two images, which is further supported by the azimuthally averaged power spectra in Fig.~\ref{fig:RC_crop}f. These power spectra were calculated in a global region in both DF images which encased the entire RC but not the blu tack that supports the sample (the larger yellow box in Fig.~\ref{fig:RC_soln}f). From these power spectra, it is evident that there is a decrease in noise, shown by the reduction in high spatial frequency components (lower noise floor), and an increase in spatial resolution, shown by the higher spatial-frequency position of the power-spectrum knees (denoted by dashed vertical lines), when the DF is considered to be rapidly varying. Although the visibility of the RC's pip and the pericarp is low in both reconstructions of the attenuation term, Figs.~\ref{fig:RC_crop}b and \ref{fig:RC_crop}d, the subject-contrast appears higher when using the previous MIST approach (when the images are shown on the same greyscale range). However, when the greyscale range is optimised independently for each attenuation term reconstruction, as shown in the magnified regions for each reconstruction, the images look almost identical. 
\begin{figure}[h]
    \centering
    \includegraphics[width=\linewidth]{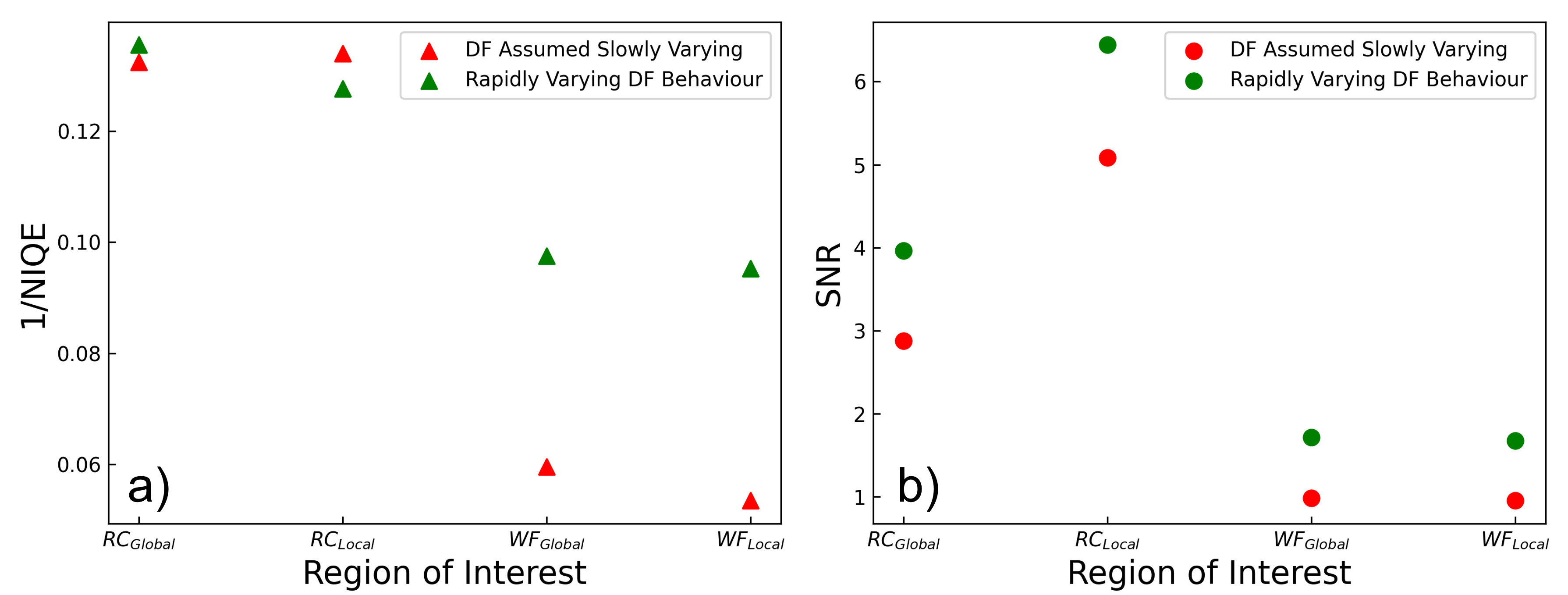}
    \caption{\textit{Image quality metrics, a) reciprocal of the NIQE and b) SNR, of the reconstructed effective diffusion coefficients using Alloo \textit{et al.}'s\cite{alloo2022dark} approach (denoted by the red ``DF Assumed Slowly Varying'' label), and that presented within this study (denoted by the green ``Rapidly Varying DF Behaviour'' label). Subscript ``\textit{Global}'' and ``\textit{Local}'' denote the region of interest used for calculations; these are indicated in Fig.~\ref{fig:WFSolutions}f and Fig.~\ref{fig:RC_soln}f for the wattle flower (WF) and red currant (RC), respectively.}}
    \label{fig:graphs}
\end{figure}

\section*{Image quality of reconstructed effective diffusion coefficients}

To quantitatively compare the WF's and RC's reconstructed DF signals using the present approach and those calculated using the DF slowly-varying attenuating-object approach presented in Alloo \textit{et al.}\cite{alloo2022dark}, we investigated the SNR and NIQE\cite{mittal2012making} for distinct regions. Two regions in each of the reconstructed DF signals were investigated for each sample; the so-called \textit{Global} region which contained the entire sample, and a \textit{Local} region which encompassed only critical structural features of the sample. The \textit{Global} and \textit{Local} regions, denoted by the large and small yellow boxes, respectively, are indicated in Fig.~\ref{fig:WFSolutions}f and Fig.~\ref{fig:RC_soln}f, for the WF and RC, respectively. The NIQE and SNR were calculated for both of the reconstructed DF signals in each of the described regions, for each sample. Figure~\ref{fig:graphs} provides a summary of the image quality metrics using the two MIST approaches. The red markers denote the DF signal calculated using the previous approach\cite{alloo2022dark} which assumed the DF to be slowly varying, but considers X-ray attenuation, and the green markers denote the new approach which considers rapidly-varying DF behaviour but only weak X-ray attenuation. It is shown from the SNR that the new approach reconstructs superior-quality DF images across all regions in both samples. The reciprocal NIQEs show similar behaviour for the DF signals, with only one datapoint as an exception, where the local reconstruction of the red current is perceived as ``better'' when the DF is assumed to be slowly varying. This scoring can be explained by the previously-mentioned point describing how the NIQE interprets the residual PC as a sharper edge, due to its increased contrast, when in reality it is incorrectly describing the DF signal. It may be argued that the overall increase in image quality shown in the remaining datapoints is only due to the filtering performed on the new DF reconstruction. However, multiple different filters (e.g., Gaussian and median) were applied to the ``DF slowly-varying reconstruction''s, for which the NIQE and SNR were calculated, and there was no such filtering that gave a comparable image quality to that obtained using the present approach.
\\\\ The described results for the WF and RC samples reveal that the present MIST approach reconstructs superior DF signals for weakly- and non-negligibly-attenuating samples, compared to our previously-published MIST approaches. We have also demonstrated that (a) for a weakly-attenuating object the proposed approach gives a qualitatively better PC reconstruction, and (b) for an attenuating sample both approaches give PC reconstructions with similar image quality. These conclusions were made based on the image quality measures of the reconstructions, rather than the quantitativeness of the reconstructions. Unsurprisingly, the quantitative difference between the reconstructed phase-shift, $\phi_{\textrm{ob}}(\textbf{r})$, from the two approaches, is larger for an attenuating object than for a weakly-attenuating object, which is reconstructed equivalently. This is exactly what is expected, based on the underlying assumptions of both theoretical formalisms. That is, the ``DF slowly-varying'' approach\cite{alloo2022dark} considers X-ray attenuation in the initial Fokker-Planck description, whereas the present approach neglects attenuation initially, before extending to the case of a weakly-attenuating material using the projection approximation. It, therefore, follows that the previous approach more accurately reconstructs the PC signal for an attenuating object, compared to the new approach. The two MIST approaches are complementary methods derived using their own distinct assumptions, therefore it is expected that their respective applicability will be sample dependent. There are regimes in which one approach is more suitable than the other. The present approach is optimal for weakly-attenuating objects (e.g., WF), however, can also be successfully applied to attenuating objects (e.g., RC).

\section*{Concluding remarks }

Within this paper, we developed and implemented a rapid deterministic approach that can reconstruct high-resolution multimodal signals of samples using SB-PCXI. The present MIST approach is not restricted to samples that have a slowly-varying DF signal, in contrast to other MIST approaches\cite{pavlov2020x,alloo2022dark,pavlov2021directional}, and instead can be used to model rapidly-varying DF behaviour. We applied the approach to two samples that differ in X-ray attenuation characteristics and compared these signals to those reconstructed using two earlier variants of MIST\cite{pavlov2020x, alloo2022dark}. Using the new approach, the SNR, spatial resolution, and perceived image quality---in the majority of local and global regions-of-interest, across the DF reconstructions for both samples---were higher.
\\\\
Multimodal X-ray imaging has already proven useful in numerous applications, and this work provides theoretical development towards reconstructing the best possible images using an SB-PCXI technique. It furthermore assists the translation of SB-PCXI into a user-friendly low-dose technique, as the proposed approach requires just 4 sets of SB-PCXI data. It is computationally efficient, requiring just 3 minutes to calculate multimodal signals for a 2100$\times$2500 pixel image using a laptop computer with an 11th Gen Intel(R) Core(TM) i7-1165G7 2.80 GHz processor and 64 GB RAM.  Moreover, the experimental setup is simple, and the SB-PCXI technique itself has low coherence requirements. 
\\\\
We anticipate that the present approach can be extended to the case of a highly-attenuating object in which the DF is spatially rapidly varying, and also to the case of rotationally-anisotropic position-dependent SAXS\cite{pavlov2021directional}. The directional DF\cite{jensen2010a,jensen2010b} approach using MIST\cite{pavlov2021directional} assumes the DF to be slowly varying. By applying the present approach to solving the directional DF inverse problem, it may be possible to optimise, by means of increased spatial resolution and SNR, the two-dimensional reconstructions, thereby furthering the future goal of MIST tensor-tomography. The present MIST approach could also be used to provide a rapidly-computed deterministic initial guess for alternative speckle-tracking approaches that solve the inverse SB-PCXI problem iteratively\cite{berujon2012two,zdora2017x,berujon2016x,berujon2017near}. Such iterative techniques---which have a broader domain of applicability because they make fewer assumptions than is the case for our work---are computationally expensive, as the multimodal signals are reconstructed with no definite initial guess. These approaches might be more computationally efficient if they had some ``good'' initial guess for the sample's multimodal signals, such as that provided by the present MIST approach. The speed of convergence of subsequent iterative refinement might significantly increase, also making these approaches more appealing for broader adoption. 

\section*{Acknowledgements}

The authors thank Sebastien Berujon, Eric Ziegler, and Emmanuel Brun for collecting and sharing the SB-PCXI data of the red currant sample collected at ESRF, published originally in Berujon \textit{et al.}\cite{berujon2016x}. The authors are grateful for the help provided by the beamline scientists, Chris Hall, Daniel Hausermann, Anton Maksimenko, and Matthew Cameron, at the Imaging and Medical beamline at the Australian Synchrotron, part of ANSTO, where the images in Figs.~\ref{fig:WFSolutions} and \ref{fig:WFSigs} were captured under proposal 18648. K. S. Morgan acknowledges support from the Australian Research Council (FT18010037). This research was supported by an AINSE Ltd. Postgraduate Research Award (PGRA) and the New Zealand Synchrotron Group Limited's capability fund grant. We acknowledge the University of Canterbury for awarding a doctoral scholarship to S. J. Alloo.

\section*{Competing interests}

The authors declare no competing interests.

\section*{Data availability}

The Python3 script, with appropriate test data, is available in the open-access repository on GitHub\cite{GITHUB}. Further experimental data are available upon reasonable request, please contact the corresponding author, S. J. Alloo. 
\bibliographystyle{unsrt} 
\bibliography{refs}

\end{document}